\documentstyle[12pt]{article}
\begin{document}
\tolerance=5000
\def\pp{{\, \mid \hskip -1.5mm =}}
\def\cL{{\cal L}}
\def\be{\begin{equation}}
\def\ee{\end{equation}}
\def\bea{\begin{eqnarray}}
\def\eea{\end{eqnarray}}
\def\tr{{\rm tr}\, }
\def\nn{\nonumber \\}
\def\e{{\rm e}}
\def\D{{D \hskip -3mm /\,}}

\def\SEH{S_{\rm EH}}
\def\SGH{S_{\rm GH}}
\def\AdS5{{{\rm AdS}_5}}
\def\S4{{{\rm S}_4}}
\def\gfv{{g_{(5)}}}
\def\gfr{{g_{(4)}}}
\def\SC{{S_{\rm C}}}
\def\RH{{R_{\rm H}}}
\def\Re{{\rm Re}}

\def\wlBox{\mbox{
\raisebox{0.3cm}{$\widetilde{\mbox{\raisebox{-0.3cm}\fbox{\ }}}$}}}
\def\htBox{\mbox{
\raisebox{0.1cm}{$\hat{\mbox{\raisebox{-0.1cm}{$\Box$}}}$}}}

\  \hfill
\begin{minipage}{3.5cm}
April 2001 \\
\end{minipage}

\begin{center}
{\large\bf Quantum bounds for gravitational de Sitter entropy 
 and the Cardy-Verlinde formula}

\vskip.5cm

{\sc S. Nojiri}$^{\heartsuit}$\footnote{nojiri@cc.nda.ac.jp}, 
{\sc O. Obregon}$^\spadesuit$\footnote{octavio@ifug3.ugto.mx}, 
{\sc S. D. Odintsov}$^{\spadesuit\clubsuit}$\footnote{
odintsov@ifug5.ugto.mx},
{\sc H. Quevedo}$^\diamondsuit$\footnote{quevedo@nuclecu.unam.mx}
and {\sc M.P. Ryan}$^\diamondsuit$\footnote{ryan@nuclecu.unam.mx}

\vskip.5cm

{\sl $^\heartsuit$Department of Applied Physics \\
National Defence Academy,
Hashirimizu Yokosuka 239-8686, JAPAN}

\vskip.5cm

{\sl $^\spadesuit$Instituto de Fisica de la Universidad de Guanajuato,
Lomas del Bosque 103, Apdo. Postal E-143, 
37150 Leon,Gto., MEXICO}

\vskip.5cm

{\sl $^\clubsuit$Tomsk State Pedagogical University, 
634041 Tomsk, RUSSIA}

\vskip.5cm
{\sl $^\diamondsuit$Instituto de Ciencias Nucleares, Universidad Nacional
Aut\'onoma de M\'exico, A. Postal 70-543, M\'exico 04510 D.F., MEXICO}

\vskip1cm

{\bf ABSTRACT}
\vskip.5cm

\end{center}

We analyze different types of quantum corrections to 
the Cardy-Verlinde entropy formula 
in a Friedmann-Robertson-Walker universe and in an 
(anti)-de Sitter space. In all cases
we show that quantum corrections can be represented 
by an effective cosmological 
constant which is then used to redefine the parameters 
entering the Cardy-Verlinde
formula so that it becomes valid also with quantum 
corrections, a fact that we
interpret as a further indication of its universality. 
A proposed relation between Cardy-Verlinde formula and 
the ADM Hamiltonian constraint is given.

\newpage

\noindent
\section{Introduction}

In the recent seminal paper by Verlinde \cite{EV} a very interesting 
approach to studying the holographic principle in a 
Friedmann-Robertson-Walker (FRW) universe filled with CFT has 
been developed. Using the dual AdS-description
\cite{AdS} a remarkable relation between entropy (energy) 
of the CFT and the equations of motion of the FRW-universe has 
been found.
An equation describing the entropy bounds during the evolution 
has been obtained \cite{EV}.
This equation has been presented in the form of the Cardy formula \cite{Cardy}
(the entropy formula for 2d CFT) which has subsequently been called
the Cardy-Verlinde formula. This approach has been further discussed 
in Refs.\cite{others,NO,WAS,SV}. In particular, 
the Cardy-Verlinde formula has been
generalized to the case of non-zero cosmological constant \cite{WAS}.
The interpretation of entropy from a brane-world perspective was  
given in Refs. \cite{SV},
and the quantum corrections to entropy from 4d QFT have been 
discussed in Ref.\cite{NO}. 

In the present work we will further discuss the entropy bounds 
and the Cardy-Verlinde formula in de Sitter (Anti-de Sitter) space, 
taking into account quantum corrections. We present the analogous model in
which quantum corrections are included as an effective cosmological 
constant. This gives us a way to formulate the quantum-corrected 
entropy bounds and the Cardy-Verlinde formula in a universal 
way (with ``renormalized'' 
parameters) where the formula takes its classical form \cite{EV}. 
As an explicit example (Anti)-de Sitter Universe is considered. 
The simple way 
to formulate quantum-corrected Cardy-Verlinde formula in a dilatonic 
Anti-de Sitter space is also outlined. We  briefly 
discuss the possibility of considering and generalizing 
the original Verlinde formula 
as an energy equation following from the ADM action.  

\ 

\noindent
\section{Quantum corrections in a FRW-universe}

Let us consider the FRW-universe equation of motion 
with quantum corrections (taking into account 
conformal-anomaly-induced 
effective action). In other words,
we suppose that the universe is filled with conformal matter 
which gives a contribution to the classical stress-energy tensor,
as well as a quantum contribution. Gravity also
makes a quantum contribution to the effective equation of motion.
 As it has been shown 
in \cite{NO}, the FRW-equation has the form:
\bea
\label{od1}
H^2&=& - {1 \over a^2} + {8\pi G \over 3}{E \over V} \nn
&& + {8\pi G \over 3}\left[
 - b'\left(4H H_{,tt} + 12 H_{,t} H^2 - 2H_{,t}^2 + 6H^4 
+ {8 \over a^2} H^2\right) \right.\nn
&& +{1 \over 12}\left\{b'' + {2 \over 3}\left(b+b'\right)
\right\} \nn
&& \quad \times \left(- 36 H_{,t}^2 + 216 H_{,t} H^2 
+ 72H H_{,tt} - {72 \over a^2}H^2+{36 \over a^4} \right) \nn
&& \left. + {\tilde a \over a^4}\right] \ ,
\eea
where $V$ is the spatial volume of the universe, 
$\tilde a=-8b'$ (a normalization choice \cite{NO}), $b''=0$ and 
\bea
\label{Ivii}
b&=&{N +6N_{1/2}+12N_1 - 8N_{\rm HD} + 611 N_2 
+ 796 N_W \over 120(4\pi)^2}\nn 
b'&=&-{N+11N_{1/2}+62N_1 -28 N_{\rm HD} + 1411 N_2 
+ 1566 N_W \over 360(4\pi)^2}\ .
\eea
Here $N$, $N_{1/2}$, $N_1$, $N_{\rm HD}$ 
are the number of scalars, (Dirac) spinors, vectors and higher 
derivative conformal scalars which are present in QFT. 
The quantity $N_2$ denotes the contribution to conformal 
anomaly from a spin-2 field 
(Einstein gravity) and $N_W$ the contribution from 
higher-derivative Weyl gravity. 

In the classical limit, only the first line of 
the FRW equation (\ref{od1}) remains 
and we obtain Einstein's dynamics with no cosmological constant. 
However, as we 
will see in a moment, the quantum corrections play 
the role of an effective cosmological constant.
 
In the absence of classical matter energy ($E=0$), 
the general FRW equation 
allows the quantum-induced de Sitter space solution 
\cite{S}:
\be
\label{od2}
a(t)=A\cosh Bt\ ,\quad ds^2= dt^2 
+ A^2 \cosh^2{t \over A}d\Omega_3^2\ ,
\ee
where $A$ is a constant and
$B^2={1 \over A^2}= - {1 \over 16\pi Gb'}$. 
On the other hand, in classical gravity with cosmological 
constant
\be
\label{od3}
H^2={8\pi G \over 3}{E \over V} 
+ {\Lambda \over 3} - {1 \over a^2} \ ,
\ee
and $E=0$, exactly the same solution (\ref{od2}) exists 
for $A^2=3/\Lambda$.
The main consecuence of this result is that 
one can consider a simplified model described by Eq.(\ref{od3})
 where the effective cosmological 
constant is defined by $\Lambda_{\rm eff}={3 \over A^2}
= - {3 \over 16\pi Gb'}$, and the quantum effects are 
encoded in the definition of $\Lambda_{\rm eff}$. 
In such a de Sitter universe we can also add classical energy.

Now one can refer to Ref.\cite{WAS}, where Verlinde's 
work \cite{EV} has been extended to include the presence of 
a cosmological constant. We define the 
Bekenstein entropy 
$S_B={2 \over 3}\pi Ea$ (the corresponding bound is valid for 
systems with limited self-gravity $Ha<1$) and the Hubble entropy 
$S_H=HV/2G$ (the corresponding bound is valid for 
strongly self-gravitating systems, $Ha>1$). 

Let us consider a black hole of the size of the universe 
described by a de Sitter space 
with metric
\be
\label{od5}
ds^2 = - f(r)dt^2 + f^{-1}(r) dr^2 + r^2 d\Omega_2^2 \ ,
\ee
where $f(r)=1 - {2MG \over r} - {r^2 \Lambda_{\rm eff} 
\over 3}$. For the maximal value of the BH mass $M$ one has the 
Nariai BH with cosmological horizon radius 
$r_c=a_c=\sqrt{{1 \over \Lambda_{\rm eff}}}$. This is the largest BH 
which may be formed in a de Sitter universe. 

Using Eq.(\ref{od3}) and supposing that the energy sufficient 
to create a SdS BH of the size of the universe is provided when 
the universe goes from a weakly to a strongly self-gravitating
phase (i.e. $Ha_c={H \over \sqrt{\Lambda_{\rm eff}}}
=\sqrt{c}$ with $c$ being 
a constant of order one), one finds 
\be
\label{od6}
H^2={8\pi G \over 3}{E_{BH} \over V} + {\Lambda_{\rm eff} \over 3} 
- \Lambda_{\rm eff}\ ,
\ee
that is 
\be
\label{od7}
E_{BH}={3\Lambda_{\rm eff} V \over 8\pi G}
\left(c+{2 \over 3}\right)\qquad {\rm or} \qquad 
S_{BH}={3\left(c+ {2 \over 3}\right)V
\sqrt{\Lambda_eff} \over 8 G} \ .
\ee

Keeping $c$ explicitly, when 
$a=a_c={1 \over \sqrt{\Lambda_{\rm eff}}}$ and 
$\Lambda=\Lambda_{\rm eff}$, one can rewrite the 
FRW equation (\ref{od3}) as the following relation between 
$S_B$, $S_H$ and $S_{BH}$  
\be
\label{od8}
S_H^2={8 \over 3\left(c + {2 \over 3}\right)}S_B S_{BH}
 - {1 \over 6}\left\{{8 \over 3\left(c + {2 \over 3}\right)}
\right\}^2 S_{BH}^2 \ .
\ee
This expression represents the quantum generalization of 
the Verlinde formula 
relating the Bekenstein, Hubble and Bekenstein-Hawking 
entropies throughout the evolution of
the universe. For $c=1$ and a classical
cosmological constant this equation was obtained in Ref.\cite{WAS}.
For $c=2/3$ and zero cosmological constant it coincides
with the original Verlinde formula \cite{EV}. Despite the fact 
that we began from  classical gravity without a cosmological constant,
the quantum effects change the dynamical structure in 
such a way that the universe 
developes an effective cosmological constant.

The above equation can be written in terms of $E$ and $E_{BH}$ as
\be
\label{od9}
S_H=\pi\sqrt{- {16\pi Gb' \over 3}}
\sqrt{{8 \over 3\left(c + {2 \over 3}\right)}E_{BH}
\left\{{2 \over 3}E 
 - {1 \over 6}{8 \over 3\left(c + {2 \over 3}\right)}
E_{BH} \right\}} \ .
\ee
If we now identify the Virasoro operator $L_0$ and the
central charge $c_{\rm CFT}$ as
\be
\label{od10}
L_0={1 \over 3}E\sqrt{-8\pi Gb'}\ ,\quad
{c_{\rm CFT} \over 2}
={8 \over 3\left(c + {2 \over 3}\right)}E_{BH}
\sqrt{-8\pi Gb'}\ ,
\ee
we can identify Eq.(\ref{od9}) as the Cardy 
formula \cite{Cardy}
\be
\label{od11}
S_H=2\pi \sqrt{{c_{\rm CFT} \over 6}
\left(L_0 - {c_{\rm CFT} \over 24}\right)}\ .
\ee
Here we have included the factor $Gb'$ in the definitions 
of $L_0$ and $c_{\rm CFT}$. Since the radius $a_c$ of the 
cosmological horizon is given by 
$a_c^2= - {16\pi Gb' \over 3}$, we can 
rewrite (\ref{od10}) as
\be
\label{od12}
L_0={1 \over \sqrt 6}Ea_c\ ,\quad
{c_{\rm CFT} \over 2}
={8 \over \sqrt 6\left(c + {2 \over 3}\right)}E_{BH}a_c\ .
\ee
Thus, we are taking into account the quantum corrections 
of the Verlinde-Cardy entropy 
formula explicitly by just redefining the Virasoro operator 
and the central charge. 
This redefinition includes only a multiplicative constant 
term and, therefore, can be 
considered as a ``renormalization'' of the quantities entering 
the Cardy formula. 
This is a further indication of the universality of Cardy's formula. 
It is also remarkable that 4d quantum dynamics 
appears in 2d quantum dynamics (the Cardy formula) via 
the corresponding
renormalization of the Virasoro operator and the central charge.

Finally, let us consider a radiation-dominated 
universe where \cite{EV}
\be
\label{odq2}
{GE \over V}\gg \Lambda_{\rm eff}\ .
\ee
If we now define the Bekenstein entropy $S_B$, the 
Bekenstein-Hawking entropy $S_{BH}$, the Hubble 
entropy $S_H$ and the quantum contribution to 
the entropy $S_{QC}$ by
\be
\label{odq3}
S_B\equiv {2\pi \over 3}Ea\ ,\quad 
S_{BH}\equiv {V \over 2Ga}\ ,\quad 
S_H\equiv {HV \over 2G}\ , \quad
S_{QC} \equiv {V \over 2G} \sqrt{ {\Lambda_{\rm eff}\over 3} }\ ,
\ee
then the effective FRW equation
\be
\label{odq1}
H^2={8\pi G \over 3}{E \over V} 
+ {\Lambda_{\rm eff} \over 3} - {1 \over a^2} 
\ee
can be written as 
\be
\label{odq4}
{S_H^2\over S_B^2} + \left(1 - {S_{BH}\over S_B}\right)^2
 - {S_{QC}^2 \over S_B^2} 
= 1 \ .
\ee
In the classical theory ($S_{QC}=0$), we obtain the original Verlinde 
formula. Since in a radiative universe ($E\propto a$) the Bekenstein 
entropy is constant, Eq.(\ref{odq4}) represents a hyperboloid with 
coordinates $S_H$, $S_B - S_{BH}$ and $S_{QC}$. In this graphical 
representation of the entropy bounds, any section 
$S_{QC}=const$ corresponds 
to a circle of radius $R_S=\sqrt{S_B^2 + S_{QC}^2}$ 
which changes as the 
volume of the universe ($S_{QC}\propto V$). Clearly, this dependence 
of $R_S$ is valid only within the range in which our 
quantum approximation
holds. For instance, in the case of an expanding universe 
the radius $R_S$
will reach its maximum value when the quantum corrections 
become incompatible
with the method applied for their derivation.   

\ 

\noindent
\section{Quantum corrections in an AdS space}

In this section we will analyze the case of classical gravity with a 
(small) negative cosmological constant, for which the action is 
given by
\be
\label{br1}
S_{\rm cl}={1\over \kappa^2}\int d^4\!x\,\sqrt{-g}\,
(R+6\Lambda)\ ,
\ee
were $\kappa^2=16\pi G$. 
Assuming that the spacetime metric is AdS$_4$
\be
\label{br2}
ds^2=\e^{2\sigma(y)}\left(-dt^2 + \left(dx^1\right)^2
+ \left(dx^2\right)^2 + dy^2\right) \ ,  
\ee
the action (\ref{br1}) reduces to
\be
\label{br3}
S_{\rm cl}=-{1\over \kappa^2}\int d^4\!x\,\e^{4\sigma}
(-6\e^{-2\sigma}((\sigma')^2+(\sigma''))+6\Lambda)\ ,
\ee
where $\sigma'={d\sigma \over dy}$. For the analysis of quantum 
corrections we add to the classical action (\ref{br1}) or (\ref{br3}) 
the trace-anomaly-induced action
\bea
\label{br4}
W&=&\int d^4\!x{\left[2b'\sigma\Box^2\sigma
-2(b+b')(\Box\sigma+\eta^{\mu\nu}
(\partial_{\mu}\sigma)(\partial_{\nu}\sigma))^2\right]} \nn
&=&V_3\int d y{\left[2b'\sigma\sigma''''
-2(b+b')(\sigma''+(\sigma')^2)^2\right]}\ .
\eea
Here $V_3$ is the space-like volume. Then by variation with 
respect to $\sigma$, we obtain the field equation 
\bea
\label{br5}
{a''''\over a}-{4a'a'''\over a^2}
-{3(a'')^2\over a^2}
+\left(6-6{b'\over b}\right){a''(a')^2\over a^3} && \nn
+{6b'(a')^4\over ba^4}
-{a\over 4b\kappa^2}\left(-12a''-24\Lambda a^3\right)&=&0\ ,
\eea
with $a=\e^\sigma$. By assuming $a={\tilde c \over y}$, 
 one obtains the following algebraic equation from 
(\ref{br5}) (we take $\Lambda<0$)
\be
\label{br6}
\kappa^2 b'-\tilde c^2-\Lambda c^4=0\;,
\ee
which has the solutions:
\be
\label{br7}
{\tilde c_1}^2=-{1\over 2\Lambda}\left(1+\sqrt{1
+4\kappa^2 b'\Lambda}\right)
\ee
and
\be
\label{br8}
{\tilde c_2}^2=-{1\over 2\Lambda}\left(1-\sqrt{1
+4\kappa^2 b'\Lambda}\right).
\ee
The first solution corresponds to the quantum corrected 
anti-de Sitter universe. Here, starting from some bare
(even very small!) negative cosmological constant, we get
an anti-de Sitter universe with a smaller cosmological constant 
due to quantum corrections. In other words, quantum corrections 
alone cannot create a 4d anti-de Sitter universe (unlike the case of 
the de Sitter universe). Rather, the quantum annihilation 
of the AdS universe may occur \cite{BO}. 

 From Eq.(\ref{br7}) one finds that the effective cosmological 
constant $\Lambda_{\rm eff}$ is given by 
\be
\label{br9}
\Lambda_{\rm eff}={2\Lambda \over 1+\sqrt{1
+4\kappa^2 b'\Lambda}}\ .
\ee
for the AdS case. Then, as in the previous section, we can 
consider an effective FRW equation,
\be
\label{br10}
H^2={8\pi G \over 3}{E \over V} 
+ {\Lambda_{\rm eff} \over 3} - {1 \over a^2} \ .
\ee
Note that $\Lambda_{\rm eff}$ is negative, which differs
from the previous section. If we assume the metric 
to be of the black hole type, as in (\ref{od5}), we have the 
following AdS black hole solution:
\be
\label{br11}
f(r)= 1 - {2MG \over r} + {r^2 \over r_0^2}\ ,
\quad r_0^2={3 \over \Lambda_{\rm eff}}\ .
\ee 
For a large black hole ($r_0\ll r 
< 2MG$), we find that the black hole radius $r_+$ (obtained from 
the condition $f(r_+)=0$) is given by
\be
\label{br12}
r_+=\left({6MG \over \Lambda_{\rm eff}}\right)^{1 \over 3} 
 - {1 \over \left(6GM\Lambda_{\rm eff}\right)^{1 \over 3}}\ ,
\ee 
or equivalently 
\be
\label{br13}
\Lambda_{\rm eff}={6GM \over r_+^3} - {3 \over r_+^2}\ .
\ee
As in \cite{WAS}, one gets the following 
expression for the Hubble bound
\be
\label{br14}
S_H^2= {8\pi^2 ER^3 \over 9G} - {4\pi ^2 R^4 \over 27G^2}\ ,
\ee
where $R$ is the universe size that in this case is given by $R=r_+$. 
We also choose $E=M$. If we then
define the Bekenstein-Hawking entropy as 
$S_{BH}={\pi R^2 \over G}$ and the Bekenstein entropy 
as $S_B=2\pi ER$, we get
\be
\label{br15}
S_H^2 = {4 \over 9}S_{BH}\left(S_B- {1 \over 3}S_{BH}\right)\ .
\ee
Defining the Virasoro operator as 
$L_0=3ER$ and the central charge as $c=24E_{BH}R$, we have
\be
\label{br16}
S_H={2\pi \over 3\sqrt{3}}\sqrt{{c \over 6}
\left(L_0 - {c \over 24}\right)}\ .
\ee
Thus, we have again arrived at the quantum-corrected Cardy-Verlinde 
formula which is exactly the same as in Ref.\cite{WAS}.
However, now the quantum corrections are taken into account
in the effective cosmological constant. As in the previous case, 
the parameters entering the Cardy formula have been 
``renormalized" in order to include 
the quantum corrections. 

When quantum effects are small, we have
\be
\label{btq1}
\Lambda_{\rm eff}=\Lambda\left(1 - k^2 b'\Lambda\right)\ 
\ee
from Eq.(\ref{br9}), and 
\be
\label{btq2}
R=r_+=R^c \left(1 + {1 \over 3}k^2 b'\Lambda\right)\ 
\ee
from Eq.(\ref{br12}). Here $R^c$ is the black hole radius
without quantum corrections:
\be
\label{btq3}
R^c\equiv\left({6MG \over \Lambda}\right)^{1 \over 3} 
 + {1 \over \left(6GM\Lambda\right)^{1 \over 3}}\ .
\ee 
In this case the the expressions for the entropies 
can be divided into a classical 
part and a part containing quantum corrections:
\bea
\label{btq4}
S_H^2&=&\left(S_H^c\right)^2 + 
\left( {8\pi^2 E\left(R^c\right)^3 \over 3G}
 - {16\pi ^2 \left(R^c\right)^4 \over 27G^2}\right)
{1 \over 3}k^2 b'\Lambda \nn
S_{BH}&=&S_{BH}^c + {2\pi \left(R^c\right)^2 \over G}
{1 \over 3}k^2 b'\Lambda \nn
S_{H}&=& S_H^c + {2\pi \over 3}R^ck^2 b'\Lambda \ ,
\eea
where 
\bea
\label{btq5}
\left(S_H^c\right)^2&=&
{8\pi^2 E\left(R^c\right)^3 \over 9G}
 - {4\pi ^2 \left(R^c\right)^4 \over 27G^2} \nn
S_{BH}^c&=&{\pi \left(R^c\right)^2 \over G} \nn
S_H^c&=&2\pi R^c \ .
\eea
The quantum-corrected Cardy-Verlinde formula (31) can easily 
be rewritten with the help of Eqs.(36) so that quantum 
corrections are given explicitly.
We should note, however, that the classical parts of the 
entropy satisfy the same relation as in Eq. (\ref{br15}):
\be
\label{btq6}
\left(S_H^c\right)^2 
= {4 \over 9}S_{BH}^c\left(S_B^c- {1 \over 3}S_{BH}^c\right)\ .
\ee

\ 

\noindent

\section{Quantum corrections in an AdS space with a dilatonic field}

We now consider the case where the dilaton $\phi$ appears in the 
conformal anomaly. For simplicity we consider an ${\cal N}=4$ 
$SU(N)$ super YM theory. The one-loop effective 
action is given by 
\be
\label{br17}
\Gamma=V_3\int d\eta \left[2b'\sigma\sigma''''
-2(b+b')(\sigma''+(\sigma')^2)
+(C\sigma+A)\Re(\phi^*\phi'''')\right]\ .
\ee
Here $A$ is a constant depending on the regularization and 
\be
\label{br18}
C={N^2-1\over(4\pi)^2}\ .
\ee
In Eq. (\ref{br17}) we assume that the spacetime metric is given by 
(\ref{br2}) and that $\phi$ also depends only on the radial 
coordinate $y$. The prime means derivative with respect to $y$. 
In Eq. (\ref{br17}) $\phi$ can be complex,
\be
\label{br19}
\phi=\chi+i\e^{-\varphi}\ ,
\ee
where $\varphi$ is the dilaton and $\chi$ is 
the R-R scalar (axion) of type IIB supergravity.
We will consider the simple case where $\chi=0$ and 
the kinetic term for the dilaton $\varphi$ in the classical 
action is absent. We also choose the regularization-dependent 
parameter $A$ to be zero. The effective equations 
of motion then take the following form:
\bea
\label{br20}
{a''''\over a}-{4a'a'''\over a^2}-{3a''^2\over a^2}
+{6a''a'^2\over a^3}\left(1-{b'\over b}\right) && \nn
+{6b'a'^4\over ba^4}+{3aa''\over\kappa^2 b}
-{C\over 4b}\,\varphi\varphi''''&=&0\ ,
\nonumber\\
\ln a\;\varphi''''+(\ln a\,\varphi)''''&=&0\ .
\eea
We now make the following change of variables:
\be
\label{br21}
dz=a(y)\, dy\;.
\ee
In terms of $z$ the first of
Equations (\ref{br20}) can be rewritten as
\bea
\label{br22}
a^2\,\mathop{a}^{....}+3a \dot{a}\,\mathop{a}^{...}
+a\ddot{a}^2-\left(5+{6b'\over b}\right)\dot{a}^2\ddot{a} && \nn
+{3\over\kappa^2 b}\left(a^2\ddot{a}+a\dot{a}^2\right)
-{C\varphi\,Y{[\varphi,a]}\over 4b}&=&0\ .
\eea
Here $\dot{a}={da \over dz}$ and $Y[\varphi,a]$ is given
in \cite{BO}, 
\begin{eqnarray}
\label{br22b}
Y[\phi,a]&=	& a^3 \stackrel{....}{\phi}+6a^2 \dot{a}
\stackrel{...}{\phi} \\
&& + 4a^2\ddot{a}\ddot{\phi}
+7a \dot{a}^2\ddot{\phi}+4a \dot{a}\ddot{a}\dot{\phi}
+a^2\stackrel{...}{a}\dot{\phi} +\dot{a}^3 \dot{\phi}\ .
\end{eqnarray} 
The second of Eqs. (\ref{br20}) (in terms of $z$) is also given
in \cite{BO} (Eq.(10)). We now 
search for special solutions of the form
\be
\label{br23}
a(z)\simeq a_0\,\e^{Hz}\;,\qquad
\varphi(z)\simeq\varphi_0\,\e^{-\alpha Hz}\;,
\ee
Analyzing the second of Eqs. (\ref{br20}) 
and dropping the logarithmic term 
(using the same arguments as in Ref.\cite{BO}),
one arrives at the solution:
\be
\label{br24}
\varphi(z) =\varphi_1\,\e^{-{3\over 2}\,{z \over l}}+
\varphi_2\,\e^{-2.62\,{z \over l}} 
+\varphi_3\,\e^{-0.38\,{z \over l}}\;,
\ee
where $\varphi_1$, $\varphi_2$, and $\varphi_3$ are constants.
Substituting the particular solution
$\varphi(z)=\varphi_0\,\e^{-\alpha {z \over l}}$
into Eq.(\ref{br22}), one obtains:
\be
\label{br25}
{1 \over l^2}\simeq{1\over\kappa^2}
\left[b'+{C\over 24}\,{\varphi_0}^2
\left(\alpha^4-6\alpha^3+11\alpha^2-6\alpha\right)\right]^{-1}\;.
\ee
The first term in the denominator is always negative, while
the second term may be positive only at $\alpha=3/2$.
It is only for the special dilaton solution
$\varphi(z)=\varphi_1\,\e^{-{3\over 2}\,Hz}$ 
(i.e.\ $\varphi_2=\varphi_3=0$) and for the condition
${\varphi_1}^2>12$ that one obtains a positive $l^2$ and,
hence, a non-imaginary scale factor for the AdS universe.
Note that the corresponding AdS scale factor is:
\be
\label{br26}
a(y)=-{l\over y}\;.
\ee
The effective cosmological constant $\Lambda_{\rm eff}$ 
can now be identified with 
\be
\label{br27}
\Lambda_{\rm eff}=-{6 \over l^2}= -\kappa^2 
\left[b'+{C\over 24}\,{\varphi_0}^2
\left(\alpha^4-6\alpha^3+11\alpha^2-6\alpha\right)\right]\ .
\ee

As one can see, the only role of the dilaton contribution is 
to change the effective cosmological constant!
As in the previous section, we can 
consider an effective FRW equation (\ref{br10}) again and find 
the black hole solution (\ref{br11}) and finally obtain the 
Cardy-like formula (\ref{br16}) taking into account the 
dilaton contribution. 
Similarly, using results of Refs.\cite{BO} one can obtain
the effective cosmological constant for de Sitter space
with non-trivial dilaton contribution. The formulas of 
the second section may again be used in this case.

\

\noindent
\section{Discussion} 

In summary, we have discussed a simple model where
quantum-corrected entropy bounds and a quantum-corrected 
(renormalized) 
Cardy-Verlinde formula are obtained.  However, 
if we study Eq. (\ref{od1}),
we see that the parts of the equation that we have replaced by
$\Lambda_{\rm eff}$ are actually a complicated combination 
of terms that
contain up to second order time derivatives of $H$.  
This leads us to the
question of how one might handle such terms if one is 
to go beyond the
simple $\Lambda_{\rm eff}$ model.  
A possibility that merits further study
is to assume that we can construct a Hamiltonian 
formulation of the problem
which can be used to identify terms in the equivalent of 
the formula for the
Hubble parameter with the proper saturated entropy bounds.

An illustration of this idea is the ADM formulation for an isotropic
$k = +1$ model with cosmological constant filled with CFT matter.  
The ADM action for this problem is
\bea
\label{ADM}
I &=& \frac {1} {16\pi G} \int [\pi^{ij}\dot g_{ij}
 - N\sqrt{g} \left (\frac
{1} {g} \left [\pi^{ij}\pi_{ij}
 - \frac {1} {2} (\pi^k_{\, k})^2\right ]
- {^3R} \right. \nn
&&\left. + \frac {16\pi GE} {V} - 2\Lambda \right ) ]dt d^3 x \ ,
\eea
where $g_{ij}$ is the three-metric on $t$ = const. surfaces and $^3R$ is the
Ricci scalar of such a surface.  If we assume that the metric has the form
$g_{ij} = a^2 (t) \tilde g_{ij}$, where $\tilde g_{ij}$ is the metric of a
unit three-sphere, the action now becomes (we take $N = 1$ to make $t$
cosmic time)
\bea
\label{ADMC}
I = \pi \int \left[ p_a \dot a - \left ( -\frac {Gp_a^2} {24 a} -\frac {6a} {G}
+ \frac {16\pi E a^3} {V} - \frac {2\Lambda a^3} {G} \right ) \right] dt,
\eea
where we have used $\int \sqrt{\tilde g} d^3 x = 16 \pi^2$.  We have also
chosen units where $p_a$ has dimensions of ordinary momentum.  The
Hamiltonian constraint (from varying $N$) results from putting the
quantity in parenthises equal to zero,
\bea
\label{Hamcn}
-\frac {Gp_a^2} {24 a} -\frac {6 a} {G} + \frac {16\pi E a^3} {V} - \frac
{2\Lambda a^3} {G} = 0.
\eea
Each of the terms in this equation has dimensions of energy, and we can write
it as
\bea
\label{ADM3}
-E_K - E_C - E_{\Lambda} + E_M = 0,
\eea
where $-E_K$ is the kinetic energy the gravitational field (a negative
quantity, as is usual for a metric where the only time dependence is in a
conformal factor), $E_C$ is the curvature energy (also negative),
$E_{\Lambda}$ the energy due to $\Lambda$, and $E_M$ is the matter energy.

A simple way to convert this to an entropy equation would be to saturate the
Bekenstein bound for each energy, but this leads to a linear relation
that is inconsistent with the Cardy-Verlinde formula.  However, it is not
difficult to show that the gravitational kinetic energy term is
$-\frac {6H^2 a^3} {G}$, where $H$ is the Hubble parameter.  In a similar
way we can identify the sum of the curvature and $\Lambda$ energies $E_C +
E_{\Lambda}$ with $S_{BH}$ of an SdS BH and the matter energy as proportional
to $S_B S_{BH}$.  These identifications allow us to conjecture that in any
theory of gravity with a Hamiltonian formulation and a Hamiltonian
constraint associated with possible redefinitions of time, we can use the
Hamiltonian constraint rewritten in terms of energies to construct the
equivalent of the Cardy-Verlinde formula for cosmological models in the
theory by identifying the gravitational kinetic energy part with $S_H^2$,
the curvature part with $S_{BH}$, and the matter part with $S_B$.  Of course,
for higher derivative theories \footnote{Quantum effective action typically 
contains the higher derivative terms.} such as 
the one given in Eq. (\ref{od1}) the
construction of a Hamiltonian formalism is difficult, and it may not be
possible to identify the terms mentioned above in such a way as to achieve
a simple version of the Cardy-Verlinde formula, but for any cosmological
model in ordinary gravity (extension of the ADM formalism to $n+1$ dimensions
is straightforward), one should be able to identify the Cardy-Verlinde
formula in the way we have outlined above.  Notice that for models with
$t$ = constant surfaces that are not conformal to an isotropic metric,
the $E_K$ term may be positive, but it can be decomposed into a negative and a
positive part, and one must be careful in identifying the new version of
$S_H$.  The Bianchi type I (Kasner) models should be a good testing ground
for our conjecture in this case.

\section*{Acknowledgments} 

The work by O.O.  was supported in part by CONACyT grant 28454E 
and that by S.D.O. was supported in part by CONACyT (CP, Ref.990356 and grant 28454E) and in part by GCFS Grant E00-3.3-461.
The work  by H.Q.  was supported in part  by DGAPA-UNAM, grant Nr. 981212.

\end{document}